\documentstyle[12pt]{article}

\begin{document}
\begin{titlepage}

\title{The gravitational energy-momentum tensor and the
gravitational pressure}

\author{J. W. Maluf$\,^{*}$\\
Instituto de F\'{\i}sica, \\
Universidade de Bras\'{\i}lia\\
C. P. 04385 \\
70.919-970 Bras\'{\i}lia DF, Brazil\\}
\date{}
\maketitle

\begin{abstract}
In the framework of the teleparallel equivalent of general relativity
it is possible to establish the energy-momentum tensor of the 
gravitational field. This tensor has the 
following essential properties: (1) it is identified directly in 
Einstein's field equations; (2) it is conserved and traceless; (3) it 
yields expressions for the energy and momentum of the gravitational 
field; (4) is is free of second (and highest) derivatives of the field 
variables; (5) the gravitational and matter energy-momentum tensors take
place in the field equations on the same footing; (6) it is unique.
However it is not symmetric. We show that the spatial components of this 
tensor yield a consistent definition of the gravitational pressure.
\end{abstract}
\thispagestyle{empty}
\vfill
\noindent PACS numbers: 04.20.Cv, 04.20.Fy\par
\bigskip
\noindent (*) e-mail: wadih@fis.unb.br\par
\end{titlepage}
\newpage

\noindent

\section{Introduction}
A comprehensive understanding of Einstein's general relativity requires
not only the knowledge of the structure of the field equations, the
solutions and physical consequences, but also the understanding of 
properties such as the energy, momentum and angular momentum of the 
gravitational field. These properties have been addressed in the 
teleparallel equivalent of general relativity (TEGR) in the past few 
years. The TEGR \cite{Hehl,Hay,Nes,Maluf1,Obukhov}
is just a reformulation of 
Einstein's general relativity in terms of tetrad fields and the torsion 
tensor. It is not a new theory for the gravitational field, because the 
field equations for the tetrad field are precisely equivalent to 
Einstein's equations. The TEGR is just an alternative geometrical 
formulation of
general relativity, and therefore it provides an alternative insight
into the theory. Investigations on pseudo-tensors indicated that
a possible expression for the gravitational energy density would be
given in terms of second order derivatives of the metric tensor. It is
known that such expression, covariant under arbitrary changes of
coordinates, does not exist. For this reason investigations on 
quasi-local gravitational energy - the energy contained within a 
closed spacelike two-surface - have been carried out (a recent review
on the subject is given in Ref. \cite{Szabados}). 
However a covariant expression that is 
linear in the derivatives of the torsion tensor can be constructed, 
and in fact a definition for the gravitational energy-momentum has been
presented \cite{Maluf2,Maluf3} and thoroughly investigated in the
framework of the TEGR. We recall that the torsion tensor cannot be made 
to vanish at a point in space-time by means of a coordinate 
transformation. Therefore criticisms based on the Principle of 
Equivalence, which rest on the reduction of the metric tensor to the
Minkowski metric tensor at a point in space-time by means of a 
coordinate transformation, do not apply to the above mentioned 
definition. It has been argued \cite{Babak} that the fact that the
first derivatives of the metric tensor can be made to vanish not only 
along the world line of a freely falling observer, but along any
world line, and independently of whether the metric tensor obeys any 
field equations, is just a feature of differential geometry.

The quasi-local expressions for the gravitational energy-momentum are
obtained by means of two general 
procedures \cite{Szabados}: the Lagrangian 
approach (the quasi-local quantities are integrals of superpotentials
derived from the Lagrangian via a Noether-type analysis) or the 
Hamiltonian approach (the quasi-local quantities are the values of the 
Hamiltonian on the constraint surface, in the phase space, as in the 
Regge and Teiltelboim \cite{RT} analysis). However in both approaches 
the resulting expressions are not uniquely determined because the action 
can be modified by adding an (almost freely chosen) boundary term to it 
(see section 3.3.3 of Ref. \cite{Szabados}). In contrast,
in the framework of the TEGR the gravitational energy-momentum $P^a$ is
identified directly in the field equations. The Hamiltonian analysis of 
the TEGR  was crucial to identifying $P^a$. It was first observed 
\cite{Maluf1} that the Hamiltonian constraint contains a scalar density 
in the form of a total divergence. The emergence of such quantity is 
possible in theories constructed out of the torsion tensor. This scalar 
density is identified as the gravitational energy density because 
integration of the latter leads to satisfactory values of the 
grativational energy for several distinct configurations of the
gravitational field. The integral form of the Hamiltonian constraint 
equation is then interpreted as 
an equation for the gravitational energy of the type $H-E=0$. In this
formulation the time gauge condition was imposed from the outset. In the 
more general Hamiltonian formulation \cite{Maluf4} the Hamiltonian and
vector constraints also contain a total divergence, the integral form of
which leads to the definition of $P^a$.

The Lagrangian formulation of the TEGR is much simpler than its 
Hamiltonian formulation. The identification of $P^a$ in the Lagrangian
field equations leads, after very simple algebraic manipulations, to a 
continuity equation for the gravitational energy-momentum, to 
conservation laws for $P^a=(E,{\bf P})$ and consequently to a definition
of the gravitational energy-momentum flux \cite{Maluf5,Maluf6}.
It is clear that in the framework of the TEGR several issues regarding 
the gravitational energy-momentum can be addressed and investigated. 

In this article we obtain the energy-momentum tensor of the 
gravitational field directly from the Lagrangian field equations. 
This tensor yields the energy-momentum, the flux of energy-momentum and 
the stresses of the gravitational field. The immediate consequence of 
this analysis is the definition of gravitational pressure. We have 
applied this definition to the simple case of the Schwarzschild
space-time. We hope that the present definition may prove to be useful 
in analises of the thermodynamics of the gravitational field.

Notation: space-time indices $\mu, \nu, ...$ and SO(3,1)
indices $a, b, ...$ run from 0 to 3. Time and space indices are
indicated according to
$\mu=0,i,\;\;a=(0),(i)$. The tetrad field and the SO(3,1) connection 
are denoted by $e^a\,_\mu$ and $\omega_{\mu ab}$, repectively.
The flat, Minkowski space-time metric tensor raises and lowers
tetrad indices and is fixed by
$\eta_{ab}=e_{a\mu} e_{b\nu}g^{\mu\nu}= (-+++)$. The determinant of the
tetrad field is represented by $e=\det(e^a\,_\mu)$.        

\bigskip

\section{The teleparallel equivalent of general relativity}

Let us consider a four-dimensional pseudo-riemannian manifold 
endowed with a set of
tetrad fields $e^a\,_\mu$ and a SO(3,1) (spin) connection
$\omega_{\mu ab}$. These quantities define the metric tensor
$g_{\mu\nu}=e^a\,_\mu e_{a\nu}$, the torsion tensor,

\begin{equation}
T^a\,_{\mu \nu}(e,\omega)=\partial_\mu e^a\,_\nu-\partial_\nu e^a\,_\mu
+\omega_\mu\,^a\,_b\,e^b\,_\nu- \omega_\nu\,^a\,_b\,e^b\,_\mu\;,
\label{1}
\end{equation}
the curvature tensor,

\begin{equation}
R^a\,_{b\mu\nu}(\omega)=\partial_\mu \omega_\nu\,^a\,_b
-\partial_\nu \omega_\mu\,^a\,_b
+\omega_\mu\,^a\,_c\, \omega_\nu\,^c\,_b
-\omega_\nu\,^a\,_c\, \omega_\mu\,^c\,_b\,,
\label{2}
\end{equation}
and the scalar curvature,

\begin{equation}
R(e,\omega)=e^{a\mu} e^{b\nu}R_{ab\mu\nu}(\omega)\,.
\label{3}
\end{equation}

\noindent The equation that defines the torsion tensor can be solved for
$\omega_{\mu ab}$. After some manipulations it is possible to obtain the 
identity,

\begin{equation}
\omega_{\mu ab}=\;^0\omega_{\mu ab}(e) + K_{\mu ab}\,.
\label{4}
\end{equation}
where $^0\omega_{\mu ab}(e)$ is the metric compatible Levi-Civita 
connection, and 

\begin{equation}
K_{\mu ab}={1\over 2}e_a\,^\lambda e_b\,^\nu(T_{\lambda \mu\nu}+
T_{\nu\lambda\mu}+T_{\mu\lambda\nu})\,,
\label{5}
\end{equation}
is the contorsion tensor.

Substitution of the identity (4) into Eq. (3) yields an
identity that relates the scalar curvature given by Eq. (3) with the 
scalar curvature $R(^0\omega)  \equiv R(e)$
constructed out the tetrad field $e^a\,_\mu$ only,

\begin{eqnarray}
eR(e,\omega)&=&eR(e)\nonumber \\
&+&e ({1\over 4}T^{abc}T_{abc}+{1\over 2}T^{abc}T_{bac}-T^aT_a)
-2\partial_\mu(eT^\mu)\,,
\label{6}
\end{eqnarray}
where $T_a=T^b\,_{ba}=T^\mu\,_{\mu a}$ and 
$T_{abc}=e_b\,^\mu e_c\,^\nu T_{a\mu\nu}$. As usual, the tetrad field 
converts space-time into SO(3,1) indices and vice-versa.
The tensor $\Sigma^{abc}$ defined by

\begin{equation}
\Sigma^{abc}={1\over 4}(T^{abc}+T^{bac}-T^{cab})+
{1\over 2}(\eta^{ac}T^b-\eta^{ab}T^c)
\label{7}
\end{equation}
yields

\begin{equation}
\Sigma^{abc}T_{abc}=
{1\over 4}T^{abc}T_{abc}+{1\over 2}T^{abc}T_{bac}-T^aT_a\,,
\label{8}
\end{equation}
and thus we have

\begin{equation}
eR(e,\omega)=eR(e)+e\Sigma^{abc}T_{abc} -2\partial_\mu(eT^\mu)\,.
\label{9}
\end{equation}

Therefore the vanishing of the curvature tensor 
$R^a\,_{b\mu\nu}(\omega)$, and
consequently of the scalar curvature given by Eq. (3), implies the 
equivalence of the scalar curvature density $eR(e)$, which defines the 
Lagrangian density for Einstein's general relativity, with the 
quadratic combination of the torsion tensor. 

We establish the Lagrangian density $L^\prime(e,\omega,\lambda)$
for the TEGR with local SO(3,1) symmetry according to

\begin{equation}
L^\prime (e,\omega,\lambda)=-ke\Sigma^{abc}T_{abc}+
\lambda^{ab\mu\nu}R_{ab\mu\nu}(\omega)-L_M\,,
\label{10}
\end{equation}
where $k=1/(16\pi G)$ and $L_M$ is the Lagrangian density for matter 
fields. $\left\{\lambda^{ab\mu\nu} \right\}$ are Lagrange multipliers 
that ensure the vanishing of the curvature tensor 
$R_{ab\mu\nu}(\omega)$. For asymptotically flat space-times the 
variation of the scalar $\Sigma^{abc}T_{abc}$ is well defined. All 
surface terms that arise in the integration by parts vanish at spatial
infinity by requiring the usual asymptotic conditions on $e_{a\mu}$,
and therefore there is no need of addition of surface terms 
to $L^\prime$. Arbitrary variations of $L^\prime$ with respect
to $e^{a\mu}$, $\omega_{\mu ab}$ and $\lambda^{ab\mu\nu}$ yield, 
respectively \cite{Maluf1},

\begin{equation}
e_{a\lambda}e_{b\mu}D_\nu (e\Sigma^{b\lambda \nu} )-
e (\Sigma^{b\nu}\,_aT_{b\nu\mu}-
{1\over 4}e_{a\mu}T_{bcd}\Sigma^{bcd} )={1\over {4k}}eT_{a\mu}\,,
\label{11}
\end{equation}

\begin{equation}
\Sigma^{a\mu b}-\Sigma^{b\mu a}-
{1\over e}D_\nu (e\lambda^{ab\mu\nu}) =-{1\over 2}S^{\mu ab}\,,
\label{12}
\end{equation}

\begin{equation}
R_{ab\mu\nu}(\omega)=0\,.
\label{13}
\end{equation}

In Eqs. (11), (12) and (13) we have the following definitions,

\begin{eqnarray}
D_\nu(e\Sigma^{b\lambda \nu})&=&\partial(e\Sigma^{b\lambda \nu})+
e\omega_\nu\,^b\,_c \Sigma^{c\lambda \nu}\,, \nonumber \\
D_\nu(e\lambda^{ab\mu\nu})&=& \partial_\nu(e\lambda^{ab\mu\nu})+
e(\omega_\nu\,^b\,_c\lambda^{ac\mu\nu}+
\omega_\nu\,^b\,_c \lambda^{cb\mu\nu})\,, \nonumber \\
{}&&{} \nonumber \\
{{\delta L_M}\over {\delta e^{a\mu}}}&=&eT_{a\mu}\,, \\
{{\delta L_M}\over {\delta \omega_{\mu ab}}}&=&eS^{\mu ab}\,.
\label{14,15}
\end{eqnarray}
The Lagrangian density $L^\prime$ as well as the field equations 
(11), (12) and (13) are invariant under SO(3,1) and general coordinate
transformations.

In order to verify the equivalence of the field equations (11) with 
Einstein's equations we substitute $\omega_{\mu ab}$ given by Eq. (4)
into $0=R_{a\mu}(e,\omega) - {1\over 2} e_{a\mu}R(e,\omega)$ (recall
that $R_{ab\mu\nu}(\omega)=0$). After long
but otherwise simple calculations we find

\begin{equation}
e_{a\lambda}e_{b\mu}D_\nu (e\Sigma^{b\lambda \nu} )-
e (\Sigma^{b\nu}\,_aT_{b\nu\mu}-
{1\over 4}e_{a\mu}T_{bcd}\Sigma^{bcd} )=
e \left[ R_{a\mu}(e) - {1\over 2} e_{a\mu}R(e)\right]\,.
\label{16}
\end{equation}
The expression that appears on the left hand side of the expression 
above is precisely the same one on the left hand side of Eq. (11). 
Therefore this feature establishes the equivalence of the TEGR with 
Einstein's general relativity. 

Let us make two remarks regarding the above Lagrangian formulation.
First, we note that the right hand side of Eq. (16) does not depend on
$\omega_{\mu ab}$, whereas the left hand side does depend on it
(this remark was first pointed out in Ref. 
\cite{Maluf1}). Therefore the SO(3,1) connection $\omega_{\mu ab}$
plays no role in the dynamics of the tetrad field, which ultimately
establishes the space-time geometry. Second, we observe that Eq. (12)
represents 24 equations. These equations are insufficient to determine
the 36 components of the Lagrange multiplier $\lambda^{ab\mu\nu}$. Some
of the latter quantities remain undetermined in the theory. From a 
different perspective, it has been pointed out in Ref. \cite{Obukhov}
that the Lagrange multipliers may be redefined by means of a symmetry
operation, which renders an ambiguity in the determination of the
multipliers. Of course this is an unsatisfactory feature of the theory.
The theory defined without the connection $\omega_{\mu ab}$ is much 
simpler and free of this undetermination, and leads to the correct
dynamics for $e_{a\mu}$. The presence of $\omega_{\mu ab}$ is not 
mandatory for the mathematical consistency of the theory. Moreover, it 
has been shown that the 
Hamiltonian formulation of the theory constructed out of $e_{a\mu}$ 
only is consistent \cite{Maluf4}. Therefore we will not take into
account the Lagrangian density $L^\prime$, and consider instead the
TEGR determined out of $e_{a\mu}$ only. By dropping out 
$\omega_{\mu ab}$ the theory is no longer invariant under local SO(3,1)
transformations, but rather invariant under global SO(3,1) 
transformations. However, the loss of this local symmetry does not 
imply any restriction on the class of possible frames. Every tetrad
field $e_{a\mu}$ that is a solution of the theory with local SO(3,1)
symmetry is also a solution of the theory with global SO(3,1) symmetry.
The global SO(3,1) symmetry implies a rigid geometric
structure in space-time that is closer in spirit to the teleparallel
geometry, as it allows the definition of distant parallelism. Finally,
we will argue later on that the global SO(3,1) symmetry is an essential
feaure of the gravitational energy-momentum, in view of the Principle of
Equivalence.

By making $\omega_{\mu ab}=0$ the torsion tensor is simplified to

\begin{equation}
T^a\,_{\mu \nu}(e)=\partial_\mu e^a\,_\nu-\partial_\nu e^a\,_\mu\,,
\label{17}
\end{equation}
and the identity (6) is rewritten as

\begin{equation}
eR(e)= -({1\over 4}T^{abc}T_{abc}+{1\over 2}T^{abc}T_{bac}-T^aT_a)
+2\partial_\mu(eT^\mu)\,.
\label{18}
\end{equation}
Considering the same definition (7) for the tensor $\Sigma^{abc}$, we
define the Lagrangian density $L(e)$,

\begin{equation}
L(e)=-ke\Sigma^{abc}T_{abc} -L_M\,,
\label{19}
\end{equation}
constructed out of $e_{a\mu}$ and matter fields only. The field 
equations derived from arbitrary variations of $L(e)$ with respect to 
$e_{a\mu}$  are given by

\begin{equation}
e_{a\lambda}e_{b\mu}\partial_\nu (e\Sigma^{b\lambda \nu} )-
e (\Sigma^{b\nu}\,_aT_{b\nu\mu}-
{1\over 4}e_{a\mu}T_{bcd}\Sigma^{bcd} )={1\over {4k}}eT_{a\mu}\,,
\label{20}
\end{equation}
where $T_{a\mu}$ is defined by Eq. (14) and $T_{a\mu\nu}$ by Eq. (17).
The theory defined by Eq. (19)
is equivalent to Einstein's general relativity because it can be shown 
that the left hand side of the equation above can be rewritten as
${1\over 2}e\left[ R_{a\mu}(e)-{1\over 2}e_{a\mu}R(e)\right]$. In the
following sections of this article we will consider the theory 
determined by Eqs. (19) and (20).

\bigskip
\section{The gravitational energy-momentum tensor}
\bigskip

The difficulty in arriving at a consistent definition for the 
gravitational energy-momentum tensor led to investigations on
pseudo-tensors. However it has become clear that this difficulty lies
in the traditional description of the gravitational field, not in the
nature of gravity as such. This point of view is considered, for 
instance, in the interesting analysis developed in Ref. \cite{Babak},
which attempts to define the gravitational energy-momentum by means of
the field-theoretical formulation of general relativity. However, 
already in Ref. \cite{Maluf2} it was antecipated that the metrical
description of gravity is not suitable for such purpose. The geometrical
formulation of the TEGR has proven to be more adequate for addressing 
the above mentioned difficulty.

In the Hamiltonian formulation of the TEGR \cite{Maluf4} in empty 
space-time the Hamiltonian and vector 
constraints, $H_0$ and $H_i$, respectively, can be arranged in order to
determine the constraint $C^a$, $C^a=e^{a0}H_0+e^{ai}H_i+\cdots\,$, which
in turn can be written as $C^a=-\partial_i \Pi^{ai} - H^a$, where 
$\Pi^{ai}$ is the momentum canonically conjugated to $e_{ai}$, and $H^a$ is
defined as the remaining part of $C^a$. The integral form of the 
constraint equations $C^a=0$ is interpreted as an equation that defines 
the vacuum gravitational energy-momentum $P^a$ \cite{Maluf3},

\begin{equation}
P^a=-\int_V d^3x\, \partial_j \Pi^{aj}\;,
\label{21}
\end{equation}
where $\Pi^{aj}=-4ke \Sigma^{a0j}$. This definition has been applied
quite satisfactorily to several gravitational field configurations. 

Returning to the Lagrangian formulation, by properly 
identifying $\Pi^{ai}$ in the field equations (20), after some simple
algebraic manipulations  it is possible to arrive at a continuity
equation for the gravitational energy-momentum \cite{Maluf5,Maluf6},

\begin{equation}
{d \over {dt}}\biggl[ -\int_V d^3x\,\partial_j \Pi^{aj} \biggr]=
-\Phi^a_g -\Phi^a_m\;,
\label{22}
\end{equation}
where $V$ is an arbitrary volume in the three-dimensional space,

\begin{equation}
\Phi^a_g=k\oint_S dS_j\lbrack e e^{a\mu}
(4\Sigma^{bcj}T_{bc\mu}-\delta^j_\mu \Sigma^{bcd}T_{bcd})\rbrack\;,
\label{23}
\end{equation}
is the $a$ component of the gravitational energy-momentum flux, and

\begin{equation}
\Phi^a_m=\oint_S dS_j\,(ee^a\,_\mu T^{j\mu})\,,
\label{24}
\end{equation}
is the $a$ component of the matter energy-momentum flux. $S$ represents 
the spatial boundary of the volume $V$. Therefore the loss of 
gravitational energy is determined by the equation 

\begin{equation}
{{dE}\over {dt}}=-\Phi^{(0)}_g-\Phi^{(0)}_m\;.
\label{25}
\end{equation}

Let us consider the Lagrangian field equation (20) in the form

\begin{equation}
\partial_\nu(-4ke\Sigma^{a\lambda\nu})=
-k\,e\, e^{a\mu}(4\Sigma^{b\nu \lambda}T_{b\nu\mu}-
\delta^\lambda_\mu \Sigma^{bdc}T_{bcd})-e\,e^a\,_\mu T^{\lambda \mu}\;.
\label{26}
\end{equation}
The $\lambda=0$ components of this equation may be written in terms of
$\Pi^{ak}$,

\begin{equation}
\partial_k (\Pi^{ak})=
-k\,e\, e^{a\mu}(4\Sigma^{b j 0}T_{b j \mu}-
\delta^0_\mu \Sigma^{bdc}T_{bcd})-e\,e^a\,_\mu T^{0 \mu}\;.
\label{27}
\end{equation}
It is useful to define the quantity $\phi^{a\lambda}$,

\begin{equation}
\phi^{a\lambda}=
k\,e\, e^{a\mu}(4\Sigma^{b c\lambda}T_{bc\mu}-
\delta^\lambda_\mu \Sigma^{bdc}T_{bcd}) \;.
\label{28}
\end{equation}
In terms of $\phi^{a\lambda}$ Eq. (27) reads

\begin{equation}
-\partial_k \Pi^{ak}=\phi^{a0}+e\,e^a\,_\mu T^{0\mu}.
\label{29}
\end{equation}
Integration of the equation above yields

\begin{equation}
P^a = \int_V d^3x \,(\phi^{a0}+e\,e^a\,_\mu T^{0\mu})\,.
\label{30}
\end{equation}
Eq. (30) suggests that $P^a$ does indeed represent the {\it total}, 
gravitational and matter fields energy-momentum.

In view of Eqs. (22), (28) and (29) we have

\begin{equation}
{{dP^a}\over {dt}}=-
\oint_S dS_j\,\phi^{aj} - \oint_S dS_j\,(e\,e^a\,_\mu T^{j\mu})\,,
\label{31}
\end{equation}
or

\begin{equation}
{d\over {dt}} \int_V d^3x\,(\phi^{a0}+e\,e^a\,_\mu T^{0\mu})=
-\int_V d^3x\,\partial_j (\phi^{aj}+ e\,e^a\,_\mu T^{j\mu})\,.
\label{32}
\end{equation}

We define the gravitational energy-momentum tensor $t^{\lambda \mu}$
as

\begin{equation}
t^{\lambda \mu}=k(4\Sigma^{bc\lambda}T_{bc}\,^\mu-
g^{\lambda \mu}\Sigma^{bcd}T_{bcd})\,,
\label{33}
\end{equation}
and therefore

\begin{equation}
\phi^{a\lambda}=e\,e^a\,_\mu t^{\lambda \mu}\,.
\label{34}
\end{equation}
In terms of $t^{\lambda \mu}$ we have

\begin{equation}
{d\over {dt}} \int_V d^3x\,e\,e^a\,_\mu (t^{0\mu} +T^{0\mu})=
-\oint_S dS_j\,
\left[ e\,e^a\,_\mu (t^{j\mu} +T^{j\mu})\right]\,.
\label{35}
\end{equation}
The total space-time energy-momentum tensor 
$t^{\lambda \mu} + T^{\lambda \mu}$ obeys the continuity equation 
(35). Thus the total space-time energy-momentum is conserved for 
appropriate boundary conditions, if the right hand side of the equation 
above vanishes under integration on the whole three-dimensional space.
Finally, we note that the field equations (26), which are
equivalent to Einstein's equations,  may be written in a simple form as

\begin{equation}
\partial_\nu(e\Sigma^{a\lambda\nu})={1\over {4k}}
e\, e^a\,_\mu( t^{\lambda \mu} + T^{\lambda \mu})\;,
\label{36}
\end{equation}
from what follows that $\partial_\lambda
\left[e\, e^a\,_\mu( t^{\lambda \mu} + T^{\lambda \mu})\right]=0$.
Equation (30) may be alternatively given by

\begin{equation}
P^a = \int_V d^3x \,e\,e^a\,_\mu(t^{0\mu}+ T^{0\mu})\,.
\label{37}
\end{equation}
However, for practical purposes expression (21) can be handled more 
easily than Eq. (37). 

The equation above shows that in order to arrive at values for the 
energy-momentum $P^a$, the energy-momentum tensor has to be
projected on a frame. Both $P^a$ and $t^{\lambda\mu}$ are frame 
dependent. We argue that it does not make sense to require quantities
such as energy and momentum to be frame independent \cite{Maluf7}. The
perception of the energy of a particle at rest, say, depends on whether 
the observer is at rest, or is under a Lorentz boost (in flat 
space-time, taking into account the 
principle of relativity, the particle may be considered as undergoing 
a boost with respect to the observer at rest), or is 
accelerated. This issue has been addressed in Ref. \cite{Maluf7}, where
it has been shown that for a moving observer that experiences a Lorentz
boost the gravitational energy of a black hole is modified by the usual
multiplicative factor $\gamma=(1-v^2/c^2)^{-1/2}$. 
The dependence of the gravitational energy with the frame 
also has to do with the principle of equivalence. According to the
principle, an accelerated frame is locally equivalent to a rest frame,
with the addition of a certain uniform gravitational field. Therefore 
the evaluation of the gravitational field on nearby bodies clearly 
depends on the state of the observer. As a consequence, a localized form
of the gravitational energy (the gravitational energy contained in a 
finite volume of space) must also depend on the frame. We note in 
addition that quantities that are invariant under local 
(in space-{\it time}) SO(3,1) symmetry are not affected by local inertial 
(frame) effects, and since local inertial effects are equivalent to local 
gravitational effects (the equivalence holds for locally uniform 
gravitational fields, for instance), such quantities are not expected to
describe any form of localized gravitational energy.

The projection of $t^{\lambda \mu}$ on a frame, as in Eq. (37), is 
essential since it allows the conservation (continuity) equation (35),
which follows from $\partial_\lambda
\left[e\, e^a\,_\mu( t^{\lambda \mu} + T^{\lambda \mu})\right]=0$. We 
recall that in the standard formulation of general relativity the 
equation $\nabla_\mu T^{\mu\nu}=0$ for the matter energy-momentum tensor
in general does not lead to conserved quantities in the space-time 
manifold.

The issue considered in this section has been
addressed in Ref. \cite{Andrade}. By analyzing the field equations (20)
in the vacuum space-time the authors of the latter reference arrive both
at a gravitational gauge current of the Yang-Mills type, $j_a\,^\mu$, 
and at a gravitational pseudo-tensor which, according to Ref. 
\cite{Andrade}, is essentially M\o ller's pseudo-tensor 
\cite{Moller1,Moller2}. The gauge current $j_a\,^\mu$ is related
to $\phi^{a\lambda}$ given by Eq. (28), which is
ultimately related to the energy-momentum tensor $t^{\lambda \mu}$ 
according to Eq. (34). M\o ller's pseudo-tensor is given by Eq. 
(26) of Ref. \cite{Andrade}, or, alternatively by
$T_\mu\,^\nu=\partial_\lambda U_\mu\,^{\nu\lambda}$, 
where \cite{Moller2}\footnote{As explained in Ref. \cite{Moller2}, 
the latter reference presents the results of Ref. \cite{Moller1}}

$$U_\mu\,^{\nu\lambda}={1\over {8\pi G}}e
\biggl[e^{a\nu}\nabla_\mu e_a\,^\lambda+
(\delta^\nu_\mu e^{a\lambda}-\delta^\lambda_\mu e^{a\nu})
\nabla_\sigma e_a\,^\sigma\biggr]\,.$$
In the expression above the covariant derivative is constructed out
of the Christoffel symbols. 

We finally remark that an expression for the energy-momentum 
tensor of the gravitational field that shares most of the properties
of Eq. (33) has been obtained in Refs. \cite{Itin1,Itin2}. In 
particular we note that the asymmetry of $t^{\lambda\mu}$ has been 
proved for a large class of teleparallel models in Ref. \cite{Itin2}.

\bigskip
\section{The gravitational pressure}
\bigskip
The gravitational energy-momentum tensor yields expressions for 
gravitational pressures and stresses, in similarity to the 
energy-momentum tensor for the electromagnetic field, for instance.
In this section we will show that it is possible to arrive at a
consistent definition of the gravitational pressure.
Restricting the following considerations to the vacuum space-time, 
Eq. (22) is simplified to 

\begin{equation}
{{dP^a}\over {dt}}=-\Phi^a_g=-\int_V d^3x\,\partial_j\,\phi^{aj}
=-\int_V d^3x\,\partial_j(e\,e^a\,_\mu t^{j\mu})\,,
\label{38}
\end{equation}
where we have taken into account Eq. (28). By requiring the index $a$ to
be a spatial index, $a=(i)=(1),(2),(3)$, we have
\begin{eqnarray}
{{dP^{(i)}}\over {dt}}&=
&-\int_V d^3x\,\partial_j\,\phi^{(i)j} 
\nonumber \\
&=&\oint_S dS_j(-\phi^{(i)j})\,. 
\label{39}
\end{eqnarray}
In the left hand side of the equation above we have the time derivative 
of a momentum component, which has the character of force. Therefore the
density $(-\phi^{(i)j})$ can be understood as force per unit area, or 
pressure density. Specifically,  $(-\phi^{(i)j})$ can be taken as a  
force exerted in the $(i)$ direction on a unit area element whose normal
points in the $j$ direction. Of course the right hand side of Eq. (39) 
can also be considered as minus the momentum flux across a 
surface $S$. In the case of gravitational waves for instance, as 
considered in Ref. \cite{Maluf5}, the momentum flux across a surface 
parallel to the wave front has the nature of a gravitational pressure.

We will make a simple application of the concept of gravitational 
pressure to the Schwarzschild space-time. The latter is characterized
by the line element

\begin{equation}
ds^2=-e^{2\lambda}dt^2 +e^{-2\lambda}dr^2+r^2d\theta^2+
r^2\,(\sin\theta)^2\, d\phi^2\,,
\label{40}
\end{equation}
where $e^{2\lambda}=1-(2m)/r$. The set of tetrad fields adapted to an 
observer at rest with respect to the black hole is given by

\begin{equation}
e_{a\mu}=\pmatrix{-e^\lambda&0&0&0\cr
0&e^{-\lambda}\sin\theta\,\cos\phi&r\cos\theta\,\cos\phi
&-r\sin\theta\,\sin\phi\cr
0&e^{-\lambda}\sin\theta\,\sin\phi&r\cos\theta\,\sin\phi
&-r\sin\theta\,\cos\phi\cr
0&e^{-\lambda}\cos\theta&-r\sin\theta&0\cr}\,.
\label{41}
\end{equation}
It is easy to verify that in the asymptotic limit 
$r\rightarrow \infty$ the tetrad fields above in cartesian 
coordinates exhibit the asymptotic behaviour 

$$e^a\,_\mu \cong \delta^a_\mu + {1\over 2} h^a\,_\mu(1/r)\,.$$
A given gravitational field configuration described by the metric
tensor $g_{\mu\nu}$ admits an infinity of tetrad fields, related to
each other by means of a local SO(3,1) transformation. In order to 
understand how an observer is adapted to a particular set of 
tetrad fields, we consider its worldline in the space-time.
Let $x^\mu(s)$ denote the worldline $C$ of an observer,
and $u^\mu(s)=dx^\mu/ds$ its velocity along $C$. We may identify
the observer's velocity with the $a=(0)$ component of $e_a\,^\mu$,
where $e_a\,^\mu e^a\,_\nu=\delta^\mu_\nu$. Thus, 
$u^\mu(s)=e_{(0)}\,^\mu$ along $C$. The acceleration of the 
observer is given by 

$$ a^\mu= {{Du^\mu}\over{ds}}={{De_{(0)}\,^\mu }\over{ds}}=
u^\alpha \nabla_\alpha e_{(0)}\,^\mu\,.$$
The covariant derivative
is constructed out of the Christoffel symbols. We see that $e_a\,^\mu$
determines the velocity and acceleration along a worldline of an 
observer adapted to the frame.
From this perspective we conclude that a given set of 
tetrad fields, for which $e_{(0)}\,^\mu$ describes a congruence of
timelike curves, is adapted to a particular class of observers,
namely, to observers determined by the velocity field 
$u^\mu=e_{(0)}\,^\mu$, endowed with acceleration $a^\mu$. If 
$e^a\,_\mu \rightarrow \delta^a_\mu$ in the limit 
$r \rightarrow \infty$, then $e^a\,_\mu$ is adapted to stationary
observers at spacelike infinity.

Out of the set of tetrad fields given by Eq. (41)
we may calculate all compoents of the
torsion tensor $T_{a\mu\nu}$, of the tensor $\Sigma^{abc}$ and
consequently of $\phi^{(i)j}$. All these are long, but simple 
calculations. 

Let us restrict considerations to a spherical surface $S$
(a surface determined by constant $r$). Then we have 

\begin{equation}
{{dP^{(i)}}\over {dt}}=-\int_S dS_1\,\phi^{(i)1}\,,
\label{42}
\end{equation}
where we are now allowing $S$ to be an open surface. After 
some calculations we find

\begin{eqnarray}
\phi^{(1)1}&=&(\sin\theta\,\cos\phi)2k\sin\theta
\biggl[ {{2m}\over r}(1-e^\lambda)-(1-e^\lambda)^2\biggr] \nonumber \\
\phi^{(2)1}&=&(\sin\theta\,\sin\phi)2k\sin\theta
\biggl[ {{2m}\over r}(1-e^\lambda)-(1-e^\lambda)^2\biggr] \nonumber \\
\phi^{(3)1}&=&(\cos\theta) 2k\sin\theta
\biggl[ {{2m}\over r}(1-e^\lambda)-(1-e^\lambda)^2\biggr] \,.
\label{43}
\end{eqnarray}
By defining the unit vector ${\hat {\bf r}}=(\sin\theta\,\cos\phi,\,
\sin\theta\,\sin\phi,\, \cos\theta)$ we may express Eq. (42) for the 
vector quantity ${\bf P}=(P^{(1)}, P^{(2)}, P^{(3)})$,

\begin{equation}
{{d {\bf P}}\over {dt}}=-\int_S d\theta\,d\phi\,(\sin\theta)
{2k} 
\biggl[ {{2m}\over r}(1-e^\lambda)-(1-e^\lambda)^2\biggr] 
{\hat {\bf r}} \,.
\label{44}
\end{equation}
We evaluate the integral in the equation above over a small
solid angle 
$\Delta \Omega =  \sin\theta\Delta \theta \Delta \phi$ of 
constant radius $r$, and consider 
a large value of $r$, $r \gg m$, such that

$$ 1-e^\lambda \approx {m\over r}={{MG}\over {c^2 r}}\,.$$

\noindent By making the replacements $dt\rightarrow d(ct)$, 
$k={1\over {16\pi}} \rightarrow {c^3 \over {16\pi G}}$, we obtain

\begin{equation}
{{d{\bf P}} \over {dt}}=-(r^2\Delta \Omega) {{M^2 G}\over r^4} 
\,{\hat {\bf r}}\,.
\label{45}
\end{equation}
The quantity on the right hand side of the equation above,
$-(M^2 G)/ r^4$, is interpreted as the 
gravitational pressure exerted on the area element $(r^2\Delta\Omega$).
As an interesting consequence of Eq. (45), we note that the latter 
equation can be rewritten as

\begin{equation}
{d\over {dt}}\biggl( {{\bf P} \over M}\biggr)
=-{{GM}\over r^2} {\Delta \Omega}\, {\hat {\bf r}}\,.
\label{46}
\end{equation}
The left hand side of Eq. (46) may be understood as a gravitational
acceleration field that acts on the solid angle $\Delta \Omega$, at 
a radial distance $r$.

For a different choice of tetrad fields Eqs. (45) and (46) 
lead to different results. It is clear that Eq. (38) transforms 
covariantly under a global SO(3,1) transformation, and so do 
expressions (42), (43) and (44). We can further consider a set of 
tetrad fields adapted to accelerated observers (the tetrad fields
considered in Ref. \cite{Maluf7}, for instance), in which case 
expressions (45) and (46) would yield different results because 
of noninertial effects.

\bigskip
\section{Concluding remarks}
\bigskip
We have presented a very simple construction of the gravitational 
energy-momentum tensor, by just considering Einstein's equations
in the teleparallel description. The energy-momentum tensor yields
the previously obtained expression for the gravitational 
energy-momentum $P^a$. The crucial point in the present analysis, as 
we pointed out earlier, is the identification of Eq. (21) as the 
gravitational energy-momentum, which is now understood as the total,
gravitational and matter fields energy-momentum. Unlike the usual 
energy-momentum tensors in classical field theory \cite{Landau}, the 
energy-momentum tensor considered above is unique, in the sense that 
it does not allow a redefinition, i.e., the addition of extra 
quantities, because these extra quantities would violate the field 
equations. An immediate issue arises regarding the definition given by 
Eq. (33): the energy-momentum tensor $t^{\lambda\mu}$ is not symmetric.
The symmetry of energy-momentum tensors is normally related to the
conservation of angular momentum \cite{Landau}. The first term in Eq. 
(33) is asymmetric. The asymmetry of similar tensors has been 
addressed in Ref. \cite{Itin2}. It has been shown that in arbitrary
teleparallel models the gravitational energy-momentum tensor cannot be 
purely antisymmetric, and that for all viable models the antisymmetric
part is nonvanishing. The meaning of the antisymmetric components of 
$t^{\lambda\mu}$ is not yet understood, and therefore this issue,
together with the conservation of the gravitational angular momentum,
deserves further investigation.

The tetrad field $e^a\,_\mu$ in Eq. (20) has 16 components. The 6
additional components (with respect to the metric tensor $g_{\mu\nu}$
establish the reference frame of a hypothetical observer. These
components characterize the rotational and translational behaviour
of the frame (this issue has been discussed in section IV of Ref.
\cite{Maluf3}). For a given gravitational field configuration
(a black hole space-time, for instance), the observer may be either
at rest, or undergoing a Lorentz boost, or may be linearly
accelerated \cite{Maluf7}.

The definition of gravitational pressure may be useful in investigations
of the thermodynamics of the gravitational field. For a given 
gravitational field configuration we may determine the energy and 
pressure of a given space volume. Therefore the analysis of the 
thermodynamic relation $TdS=dE + pdV$ may be extended from the horizon 
area of black holes to arbitrary volumes in space. Of course it is not
straightforward to generalize the above thermodynamic relation in this 
sense, because the notion of gravitational entropy is so far connected 
with the area of the horizon of a black hole.

\end{document}